\documentclass[aps,prd,eqsecnum,onecolumn,showpacs,nofootinbib,preprintnumbers,a4paper]{revtex4}

    \newcommand*{\be}{\begin{equation}}
    \newcommand*{\ee}{\end{equation}}
    \newcommand*{\ba}{\begin{eqnarray}}
    \newcommand*{\ea}{\end{eqnarray}}

\usepackage{slashed} 
\usepackage{amssymb}
\usepackage{array} 
\usepackage{hyperref}

\begin{document}

\title{Supersymmetry Breaking and Dilaton 
Stabilization in String Gas Cosmology}

\author{Sasmita Mishra$^{1,2}$, Wei Xue$^{1)}$, 
Robert Brandenberger$^{1)}$ and Urjit Yajnik$^{2,1}$}

\affiliation{1) Physics Department, McGill University, Montreal, 
QC, H3A 2T8, Canada}

\affiliation{2) Physics Department, IIT Bombay, India}

\begin{abstract}

In this Note we study supersymmetry breaking via gaugino
condensation in string gas cosmology. We show that
the same gaugino condensate which is introduced to stabilize 
the dilaton breaks supersymmetry. We study the constraints
on the scale of supersymmetry breaking which this
mechanism leads to.

\end{abstract}

\maketitle

\section{Introduction}

String Gas Cosmology \cite{BV} (see also \cite{Perlt}
for an original reference,\cite{RHBSGrev,BattWat} for
reviews, and \cite{others} for a selection of other earlier
papers) is an approach to superstring cosmology based
on making use of degrees of freedom and symmetries which
are particular to string theory and studying their
effects in the very early universe. More specifically,
string gas cosmology rests on coupling the energy-momentum
tensor of a thermal gas of perturbative string states including
modes with momentum and winding about the extra spatial
dimensions to a background space-time and studying the
resulting dynamics. 

One of the first results which
emerges \cite{BV} is that the temperature singularity
of Standard (and also Inflationary) Cosmology is resolved:
the temperature $T$ of the string gas can never exceed a maximal
temperature $T_H$ called the Hagedorn temperature \cite{Hagedorn}.
In fact, if space is compact and isotropic, described by a radius
$R$, the the T-duality symmetry of string theory implies
\be
T(R) \, = \, T(1/R) \, ,
\ee
where in this formula the radius is expressed in units of
the string length. Thus, the evolution of the very early
universe is clearly going to be very different from what
is expected based on intuition gained from Standard and
Inflationary cosmology. In fact, it is possible that
in string theory the universe begins with a long quasi-stationary 
phase with a temperature
close to $T_H$. This conjectured phase is called the
``Hagedorn phase'' \footnote{This phase cannot be
described by either General Relativity or Dilaton Gravity 
(see \cite{RHBetal,KKLM} for discussions of this point). For
an attempt to obtain a Hagedorn phase in the context of an
effective field theory see e.g. \cite{BFK}.}. The Hagedorn
phase has a smooth transition to the expanding phase of
Standard Cosmology which is given by the decay of string
winding modes into string loops.

As was recently realized \cite{NBV} (see also \cite{BNPV2}),
thermal fluctuations in the Hagedorn phase of string gas
cosmology evolve into a scale-invariant spectrum of
cosmological perturbations at late times. Thus, string gas
cosmology provides an alternative to inflationary cosmology
in providing an origin for the observed fluctuations in the
distribution of matter and microwave radiation in the
universe. A specific prediction of string gas cosmology
with which the scenario can be distinguished from any
inflationary model \footnote{Any model based on General
Relativity as the theory of space and time and matter obeying
the usual energy conditions.} is a slightly blue spectrum
of gravitational waves \cite{BNPV1}.

In order for string gas cosmology to be consistent with
late time cosmology, the moduli describing the size
and shape of the extra dimensions of space must be
stabilized. In Heterotic string theory this is achieved
naturally by means of string states containing both
winding and momentum about the extra dimensions which
become massless at the self-dual radius (of the extra
spatial dimensions). These states fix both the size
moduli \cite{Patil2} (see also \cite{Watson1,Watson2,Patil1} for
earlier work) and shape moduli \cite{Edna} \footnote{See also
\cite{others2} for a selection of other papers on moduli 
stabilization in string gas cosmology.}.  Note that
at this stage no new extra ingredients had to be
introduced. 

In order to be phenomenologically viable, the dilaton
must also be fixed. This has recently been achieved
by making use of a nonperturbative mechanism used
frequently in different contexts: gaugino condensation.
In \cite{Danos:2008pv} it was shown that gaugino condensation
fixes the dilaton without destroying the stabilization
mechanism of the radion \footnote{See \cite{others2} for
other works on moduli stabilization in string gas cosmology.}. 

In this Note we wish to point out the the same gaugino
condensation mechanism used to fix the dilaton automatically
breaks supersymmetry. We study the breaking of
supersymmetry in detail and show that supersymmetry
breaking at both low and high scales is consistent with
dilaton and radion stabilization. Thus, it appears that
string gas cosmology is consistent with both late time
cosmology and particle phenomenology.

In the following section we briefly review string gas cosmology
and the stringy mechanism which stabilizes both the size
(radion) and shape moduli associated with the extra spatial
dimensions. In Section 3 we review the work of \cite{Danos:2008pv}
which shows how assuming gaugino condensation leads to a
mechanism to stabilize the dilaton without disrupting
the stringy stabilization of the radion. We then proceed
to demonstrate that the same gaugino condensation mechanism
breaks supersymmetry. Finally, we compute the dilaton mass,
the gravitino mass and the supersymmetry breaking scale in
terms of the constants which appear in the superpotential
of gaugino condensation and discuss phenomenological
constraints.

\section{Review of String Gas Cosmology}

In string gas cosmology \cite{BV}, matter is taken to be
a thermal gas of perturbative string states. These include
modes which contain both momentum and winding about the
extra spatial dimensions. We assume that the topology
of the internal spatial manifold is such that long-lived
winding modes exist (see e.g. \cite{Brian} for a discussion
of which orbifold compactifications admit long-lived
winding modes). 

Let us now follow the evolution of this thermal
string system as we go backwards in time when space
decreases in size and the energy density rises. Initially,
the energy density of the string gas is in the states which are
light at large radii, namely the string momentum modes. 
However, as we go backwards in time, eventually the thermal
energy will be so large that it becomes possible to excite the
string oscillatory modes. Since the number of string states
rises exponentially with energy \cite{Hagedorn}, there is
a maximal temperature which the thermal string gas can
reach, the so-called Hagedorn temperature $T_H$. Instead
of increasing the energy of the modes which are already
excited, the extra energy density which is obtained if the
volume of space decreases goes to exciting new modes.
Eventually, even string winding modes are excited.
The string gas of the very early universe is hence expected
to contain all perturbative string states, in particular
modes which wind the internal spatial dimensions.

The equations which govern the dynamics of the early
phase of string gas cosmology when the temperature
hovers close to (but below) $T_H$ are not known. They
cannot be those of General Relativity nor those
of Dilaton Gravity, since neither set of equations are
consistent with the full set of symmetries of string
theory \footnote{Dilaton gravity is consistent with the
T-duality symmetry of string theory, but for large values
of the dilaton the string states are no longer the
lightest ones and instead D-brane states become
dominant.}. However, at late times when the temperature
is significantly smaller than $T_H$, the equations
of motion which describe the background space-time
must reduce to those of Dilaton Gravity coupled to
a thermal gas of strings as the matter content.

Thus, the action which describes string gas cosmology
in ten space-time dimensions at late times is
\begin{equation}
 S \, = \, \frac{1}{k}(S_g + S_{\phi}) + S_{SG} \, ,
\label{eq:total-action}
\end{equation}
where $S_g$ is the string frame gravitational action (which
has a dilaton dependence), 
$S_{\phi}$ is the dilaton action and $S_{SG}$ is action 
for the string gas. Also,
\begin{equation}
 k \, = \, 16\pi G \, = \, \frac{16\pi}{M_{10 }^8} \, ,
\end{equation}
where $M_{10}$ is the $10D$ Planck mass.

The dilaton action is given by
\begin{equation}
 S_{\phi} \, = \, -\int d^{10}x \sqrt{-g}\left[
\frac{1}{2}\partial_{\mu} \varphi  \partial^{\mu} \varphi
+ k V(\varphi)\right] \, ,
\label{eq:dilaton-action}
\end{equation}
where $\varphi$ is the ten-dimensional dilaton. In the
absence of non-perturbative effects, the dilaton potential
$V(\varphi)$ vanishes. It is known \cite{RHBetal} that
without a potential the dilaton cannot be stabilized.
We will follow \cite{Danos:2008pv} and consider the
non-perturbative gaugino condensation mechanism which
leads to a non-vanishing dilaton potential.
 
The string gas is treated as ideal gas. Hence, the action 
can be written as (see e.g. \cite{Patil1,Patil2} for extensive
discussions)
\begin{equation}
 S_{SG} \, = \, -\int  d^{10}x\sqrt{-g}\sum_{\alpha}
\mu_{\alpha}\epsilon_{\alpha} \, ,
\label{eq:stringGas-action}
\end{equation}
where the index $\alpha$ runs over all string states,
 $\mu_{\alpha}(x,t)$ is number density of strings in the state
$\alpha$, and $\epsilon_{\alpha}(t)$ is the energy of this string 
state. Factoring out the expansion of the universe we have
\begin{equation}
 \mu_{\alpha} \, = \, \sqrt{-g_{ss}}^{-1} \mu_{0,\alpha}(t) 
\end{equation}
where $\mu_{0, \alpha)}$ is the comoving number density
and $g_{ss}$ is the determinant of the spatial part of the metric:
\begin{equation}
 \sqrt{-g} \, = \, \sqrt{-g_{ss}}\sqrt{-g_{00}} \, .
\end{equation}
With these substitutions, the string gas action becomes
\begin{equation}
 S_{SG} \, = \, - \int d^{10}x \sqrt{-g_{00}}
\sum_{\alpha}\mu_{0,\alpha}(t)\epsilon_{\alpha} \, .
\label{eq:stringGas-action-1}
\end{equation}
Since it is important to understand how size moduli stabilization in
string gas cosmology comes about, we list below the expressions
for the energy density and the pressures of the string gas in the various
spatial directions.

We will write down the expressions for the energy density and the
various pressures in the case of an anisotropic but homogeneous
background metric given by
\begin{equation}
 ds^2 \, = \, -dt^2 + a(t)^2 dx^2 + \sum_{a=1}^6 b_a(t)^2 dy_a^2 \, ,
\label{eq:metric}
\end{equation}
where $a(t)$ is the scale factor of our three-dimensional space and
$b_a(t)$ is the scale factor of the a'th direction of the internal space
(here, for simplicity, we have an internal torus in mind).

By varying the above string gas action with respect to the metric, 
the expressions for the non-vanishing components of the 
energy momentum-tensor can be derived. The contribution of the
string state $\alpha$ to the energy density is
given by
\begin{equation}
 \rho_{\alpha} \, = \, \frac{\mu_{0,\alpha}}{\epsilon_{\alpha}\sqrt{-g}}
\epsilon_{\alpha}^2 \, ,
\label{eq:rho}
\end{equation}
and that to the three (large) dimensional pressure by,
\begin{equation}
P_{\alpha}^i = \frac{\mu_{0,\alpha}}{\epsilon_{\alpha}\sqrt{-g}}
\frac{P_d^2}{3}
\label{eq:3d-Pressure}
\end{equation}
where $P_d$ is the momentum in the $d=3$ large dimensions. The
contribution of the string state $\alpha$ to the pressure in the
a'th direction of the internal compact space is given by
\begin{equation}
 P_{\alpha}^a \, = \, \frac{\mu_{0,\alpha}}{\epsilon_{\alpha}\sqrt{-g}
\alpha'}\left( \frac{n_a^2}{b_a^2} - w_a^2b_a^2\right) 
\label{eq:compactDim-pressure}
\end{equation}
where $n_a$ and $w_a$ are the momentum and winding numbers, respectively,
of the string state in the a'th direction (we are omitting the label $\alpha$
on $n$, $w$ and $N$ to keep the notation simpler). 
For the sake of completeness
we also give the expression for the energy $\epsilon_{\alpha}$ of the
state $\alpha$:
\begin{equation}
 \epsilon_{\alpha} \, = \, \frac{1}{\sqrt{\alpha'}}
\left[ \alpha' p_d^2 + b^{-2}(n,n)+
b^2(w,w)+2(n,w)+4(N-1)\right]^{1/2}
\label{eq:epsilon-alpha}
\end{equation}
where $n$ and $w$ are the momentum and winding number
vectors in the internal space, and $N$ is the oscillator level.
The parentheses in $(n,n)$, $(w,w)$ and $(n,w)$ indicate the scalar
product of the momentum and winding number vectors in the
internal dimensions.

The size moduli $b_a(t)$ and shape moduli (which we have
implicitly set to zero in the above ansatz for the metric)
must be stabilized at low energy densities in order that the 
resulting low energy effective theory be consistent with
experimental and observational constraints. The stabilization
of the size moduli can be understood by
looking at the above formula (\ref{eq:compactDim-pressure})
for the pressure in one of the compact directions: there are
special states which have both momentum and winding about
the extra dimensions for which this pressure vanishes at the
self-dual radius. If these
states dominate the string gas partition function, then 
the effective potential for the dynamics of $b_a(t)$
will have a stable minimum at a value given by the 
``self-dual radius" which in general is the string scale
\footnote{A heuristic way to understand radion stabilization is
as follows \cite{Watson1}: the winding modes counteract
an increase in the value of the radion (see Appendix A
for the definition of the radion field $\sigma(t)$ in terms
of $b(t)$), whereas the momentum
modes oppose a contraction, leaving the self-dual point
as a stable point of the dynamics.}. 
As discussed in \cite{Patil2}, in heterotic string theory
these special states are massless at the self-dual radius.
Hence, if the value of $b$ is close to the self-dual
radius, these states will dominate the string partition
function, and they will lead to radion stabilization.
They are enhanced symmetry states which trap the radion
at the fixed point \footnote{See \cite{Watson2,Silverstein}
for more general discussions of moduli trapping at
enhanced symmetry points.}. Assuming that the dilaton
is fixed, it was shown in \cite{Patil2} that the enhanced 
symmetry point is a stable fixed point for the radion,
and that the fluctuations about this enhanced symmetry
point are phenomenologically safe \footnote{Shape
moduli stabilization in the same perturbative superstring
context was studied in \cite{Edna}.}

To be phenomenologically viable, string gas cosmology
must also admit a mechanism which stabilizes the dilaton.
In \cite{Danos:2008pv}, a gaugino condensation mechanism
as a means to stabilize the dilaton was explored. By
studying the dilaton potential $V$ arising from the
superpotential of gaugino condensation, it was shown
that the dilaton can be trapped at a stable fixed point.
In fact, it was shown that dilaton stabilization by gaugino
condensation is consistent with radion stabilization by
string gases. This was shown by expanding the equations
of motion for both the radion and the dilaton about the
fixed point, and showing that the mass matrix for the
fluctuations is positive definite.

Gaugino condensation is a well-known mechanism to break
local supersymmetry. In the following, we will
explore this mechanism in the context of string gas
cosmology.

\section{Gaugino Condensation and Supersymmetry Breaking}

We now study the gaugino condensation mechanism
and the resulting breaking of supersymmetry. We
are working in the context of perturbative heterotic
$E_8 \times E_8$ string theory. Our goal is to 
find a stable state that breaks
supersymmetry but keeps the cosmological constant zero.

We consider a background with space-time $M^4 \times K$, where
$M^4$ is our four-dimensional Minkowksi space-time and $K$ 
is a six-dimensional manifold of $SU(3)$ holonomy (and needs
to be chosen such that the enhanced symmetry states which
can lead to the size and shape moduli stabilization actually exist
- for a specific example see \cite{Danos:2008pv}). This 
compactification breaks $E_8\times E_8$ to $E_6\times E_8$
\cite{heterotic}. Subsequently the $E_6$ breaks to the Standard
Model gauge group. The second $E_8$ is the hidden sector of low energy
supergravity which breaks to a subgroup $Q$. 

Let us consider the ten-dimensional supergravity Lagrangian for 
gauge fields and gravitons which is given by (following the
notation in the textbook \cite{Pol}):
\begin{equation}
 {\mathcal L} \, = \, \frac{M_{10}^8}{16\pi}R - \frac{M_{10}^8 \, e^{-\varphi}}{32\pi}|H_3|^2-
\frac{e^{\varphi/2}}{4g_{10}^2} {\rm Tr} F_{AB}\,F^{AB}
\end{equation}
where $g_{10}$ is the ten-dimensional
gauge coupling, $H_3$ is the NS-NS field and $F_{AB}$ is the gauge field
(the capitalized Latin indices run over all of the dimensions). 
The above action can be reduced to a 
four-dimensional gauge action which is given by
\begin{equation}
 {\mathcal S} \, = \, \int d^4x\, \frac{1}{4}\, 
e^{-2\varphi}b_E^6\,(M_{10}^6g_{10}^2)^{-1}
{\rm Tr} F_{\mu\nu}^2 
\label{eq:4d-action}
\end{equation}
where the Einstein frame scale factor $b_E$ of the extra dimensions yields the radion
$\sigma$ (see Appendix A).
Setting $M_p^6g_{10}^2 = 1$, the four-dimensional gauge coupling
can be read off from eq.(\ref{eq:4d-action}) as 
\begin{equation}
 g_4^2 \, = \, e^{2\varphi}b_E^{-6} \, .
\label{eq:g-phi-relation}
\end{equation}

The low energy subgroup $Q$ becomes strong
at a mass scale $\mu$ given by
\begin{equation}
\mu \, \sim \, M_{10} b_E^{-4} exp(-1/(2b_0 g_4^2)) \, ,
\label{eq:mu-g-relation}
\end{equation}
where $b_0$ is the one-loop $\beta$ function
coefficient of $Q$. In order for the ten-dimensional 
Planck mass to coincide with the four-dimensional Planck
mass, the corresponding metics are related by
\be
g_{\mu\nu}^{(10)}\, = \, b_E^{-6} g_{\mu\nu}^{(4)} \, .
\ee
The massive gauge boson mass is of order 
\be
M_{\rm GUT} \, = \, M_{10} b_E^{-4}\, .
\ee
Inserting the value of $g_4$ from Eq.(\ref{eq:g-phi-relation})
intp Eq.(\ref{eq:mu-g-relation}), the mass scale
$\mu$ depends on $\varphi$ as \cite{Dine:1985rz} (see also \cite{Derendinger})
\begin{eqnarray}
 \mu \, &\sim& \, M_{\rm GUT} exp(-1/(2 b_0 e^{2\varphi}b_E^{-6}))\\
&\sim& \,  M_{\rm GUT} exp(-1/(2b_0 e^{2\varphi})) \, ,
\label{eq:mu-phi-relation}
\end{eqnarray}
where we have used that $b_E$ can be set to $1$ if radion
stabilization occurs at the string scale (see Appendix A).
The above equation indicates that for small values of
$\varphi$, the characteristic scale of $Q$, namely $\mu$, is
much less than $M_p$.

So for this region gravity is a small correction as
compared to gauge forces of $Q$. 
Now considering the possibility of supersymmetry breaking in this
sector, there is a possibility of formation of gaugino
condensates through which supersymmetry breaking can be triggered.
Since $\mu$ is the mass scale for $Q$, gauginos of
the gauge superfield $Q$ can condense with 
\begin{equation}
 \langle {\rm Tr \chi\bar{\chi}}\rangle \, \sim \, \mu^3
\label{eq:chi-mu}
\end{equation}
from instanton effects. However it has been shown in
\cite{Veneziano:1982ah} that these gaugino condensates can not
break supersymmetry in pure gauge theories (i.e. in the case
of global supersymmetry). The situation is improved \cite{Dine:1985rz}
by coupling the gauge theory to supergravity that contains
an axion field. This axion fields arises while compactifying
ten-dimensional supergravity to four dimensions, and it is the 
Hodge dual of the field $H_3$.  

Thus, the gaugino condensation in four dimensions corresponds to
\begin{equation}
\langle {\rm Tr \chi\Gamma_{ijk}\bar{\chi}}\rangle \, = \,  
A \epsilon_{ijk}
\end{equation}
in ten dimensions, where $A$ is a complex number. 

One way to see that gaugino condensation triggers
supersymmetry breaking is to apply a
supersymmetry transformation on the supergravity multiplet. 
Under a supersymmetry transformation,
the massless fermion $\chi$ of the supergravity multiplet
transforms as 
\begin{equation}
 \delta \lambda \, \sim \, ({\rm Tr} \bar{\chi}\chi)\epsilon +
({\rm Tr} \bar{\chi}\gamma_5\chi)\gamma_5\epsilon
\end{equation}
so that if ${\rm Tr} \bar{\chi}\chi$ or ${\rm Tr} \bar{\chi}\gamma_5\chi$
is non-zero (refer to Eq.(\ref{eq:chi-mu})) then $ \delta \lambda \ne 0$,
$\chi$ is a Goldstone fermion and supersymmetry is spontaneously broken. 

The scale of supersymmetry breaking $M_s$ and the mass of the
gravitino $m_{3/2}$ will be used later. Their values in terms
of $\mu$ are given in Appendix 2. In the following we will
also need the expression for the dilaton mass. To obtain its
expression, we need to study the dilaton potential induced
by gaugino condensation.

Gaugino condensation is usually analyzed in terms of the
fields appearing in the four-dimensional effective field
theory which describes our dimensions. We must
thus pause for a moment and relate the fields in
the four-dimensional theory to those in the ten-dimensional
space. The four-dimensional dilaton $\Phi$ is given by
\be \label{rel}
\Phi \, = \, 2\varphi - 6\,ln b_E \, ,
\ee
which leads to the relationship between gauge coupling and 
dilaton field in four dimensions,
\be 
g_4^2 = e^\Phi \ .
\label{g4}
\ee

The first basic potential in a supersymmetric theory is the
K\"ahler potential $K(S, T)$. We take it to be of minimal form,
namely
\begin{equation}
K(S,T) \, = \, -ln(S+S^*)-3\, ln (T+T^*) \, ,
\label{eq:moduli-kahler}
\end{equation}
where $S$ is the dilaton-axion multiplet
\begin{equation}
S \, = \, e^{-\Phi} + ia \, ,
\end{equation}
$a$ being the axion field, and $T$ is
\be
T \, = \, b^2 + i \beta \, ,
\ee
$b$ corresponding to the radion and $\beta$ describing the flux
about the compact directions (which we set to zero
in our model).

Compactifications with gaugino condensation lead to
a superpotential of the form
\begin{equation}
W \, = \, M_{4}^3(C - A e^{-a_0S}) 
\label{eq:gaugino-condensation-superpotential}
\end{equation}
where $C, a_0$ and $A$ are constants, and $M_{4}$ is the
four dimensional Planck mass. The second term
arises from gaugino condensation, the first comes from
fluxes and a Chern-Simons term. The flux term must be
added in order to obtain vanishing potential energy
in the ground state.

The scalar potential in the four-dimensional Einstein frame is given by
\begin{equation}
V_F \, = \, \frac{1}{M_{4}^2} e^{\kappa}\left[K^{A \bar{B}}D_A W
D_{\bar{B}}\bar{W} - 3 |W|^2 \right] 
\label{eq:sugra-VF}
\end{equation}
where $A, B$ run over all moduli fields ($\Phi$ and $b$ in
our case). The K\"ahler covariant derivatives are given by
\begin{equation}
 D_A W \, = \, \partial_A W +(\partial_A K)W \, .
\label{eq:covariant-derivative}
\end{equation}
Here the superpotential is independent of the $T$ modulus field.
Thus, Eq.(\ref{eq:sugra-VF}) reduces to 
\begin{equation}
 V \, = \, \frac{1}{M_{4}^2}e^K K^{a \bar{b}}D_a W D_{\bar{b}}\bar{W} \, ,
\label{eq:reduced-sugra-VF}
\end{equation}
where $a, b$ now runs over $S$. After simplifying the
above expression the scalar potential is given by,
\begin{equation}
V \, = \, \frac{M_{4}^4}{4} b^{-6}e^{-\Phi}\left[
\frac{C^2}{4}e^{2\Phi} - AC e^{\Phi} \left(a_0+\frac{1}{2}
e^{\Phi}\right)e^{-a_0e^{-\Phi}} +
A^2\left(a_0+\frac{1}{2}e^{\Phi}\right)^2e^{-2a_0e^{-\Phi}} \right] \, .
\label{eq:scalarPotentil}
\end{equation}
Minimising the above potential and expanding around a stable
minimum $\Phi = \Phi_0$, the approximate form of the potential is given by
\begin{equation}
 V \, = \, \frac{M_{4}^4}{4} b^{-6}e^{-\Phi_0}a_0^2A^2
\left(a_0 - \frac{3}{2}e^{\Phi_0}\right)^2e^{-2a_0e^{-\Phi_0}}
(e^{-\Phi}-e^{-\Phi_0})^2 \, .
\label{eq:approx-pot}
\end{equation}
From this expression for the dilaton potential we can
compute the dilaton mass obtained if the dilaton sits
at the minimum of its potential.

For the sake of completeness we can also write down
the potential in terms of ten-dimensional fields.
Using the relation (\ref{rel}) between the four-dimensional 
and ten-dimensional dilatons and the following relation between the 
radions in the string and Einstein frames
\be
b_s \, = \, e^{\varphi/4}b_E \, ,
\ee
the approximate form of the scalar potential can be written as
\begin{equation}
 V(b,\varphi) \, = \, \frac{M_{10}^{16}\hat{V}}{4}
 e^{-\Phi_0}a_0^2A^2 \left(a_0 - \frac{3}{2}e^{\Phi_0}\right)^2e^{-2a_0e^{-\Phi_0}}
e^{-3\varphi/2}(b^6e^{-\varphi/2}-e^{-\Phi_0})^2
\label{eq:10D-scalarPot}
\end{equation}
where we have used the following relation between the four-dimensional 
Planck mass $M_{4}$ and the ten-dimensional one $M_{10}$: 
\be
M_{4}^2 \, = \, M_{10}^8\hat{V} \, .
\ee

\section{Phenomenological Considerations}

It is now important to investigate what values of the
supersymmetry breaking scale can arise from string
gas cosmology. We must consider a range of constraints.
Firstly, string gas cosmology is based on the assumption
that the string coupling constant is small, i.e. that
\be
\Phi_0 \, < \, 0 \, .
\ee
If this condition were not satisfied, then the gauge 
coupling is not weak and the string
states would not be the lightest states. In
particular, D-branes would have to be considered. On
the other hand, if the string coupling constant is
small, then D-branes will decouple early in the cosmological
evolution, leaving the perturbative string states
to dominate the late-time dynamics \cite{ABE}.

Since the dilaton potential takes the form of a perfect
square, the cosmological constant will vanish at
the minimum of the potential. The conditions for
$\Phi_0$ to be a minimum of the potential are
\be \label{cond1}
|D_T W|^2 \, = \, 3 |W|^2 \, 
\ee
and
\be \label{cond2}
|D_S W|^2 \, = \, 0 \, ,
\ee
the first of which is automatically satisfied,
leaving (\ref{cond2}) as the condition which is
to be verified. For our minimal K\"ahler potential
(\ref{eq:moduli-kahler}) and our superpotential 
(\ref{eq:gaugino-condensation-superpotential}) this
condition becomes
\be \label{cond3}
A \bigl( 2 {\rm{Re}}(S) a_0 + 1 \bigr) e^{- a_0 S} \, = \, C \, ,
\ee
where $S$ is evaluated at the minimum $\Phi_0$. 

The equation (\ref{cond3}) provides a relation between the
three constants $A, C$, and $a_0$ in the superpotential
of gaugino condensation. Note that it should not be
unexpected that the constraint of vanishing potential
at the minimum leads to a constraint on the parameters
in the underlying theory: in all of our current models of
particle physics a fine-tuning of coefficients is required
in order to eliminate what would otherwise be a large
cosmological constant.

Furthermore, there are phenomenological constraints
on the dilaton and the gravitino. They must
either be extremely light or quite heavy. If they
are very light, their decay time would be 
cosmological, and they could dominate the energy
density of the universe. This leads to the constraint
that the mass of a long lived gravitationally coupled
particle should be less than about $1$KeV \cite{overab}.
A much tighter constraint comes from fifth force
constraints and from the life-time of white dwarfs
and red giants \cite{Tsamis}. The bound is
\be
m_X \, < \, 10^{-33} {\rm{eV}} \, ,
\ee
where $X$ refers to both the gravitino and the dilaton.
This window will be uninteresting for us.

If the particles are heavy and decay on time
scales shorter than the life-time of the universe
there are also cosmological constraints. In particular,
the mass should obey \cite{deCarlos,Ellis}
\be
m_X \, > \, 10^4 {\rm{GeV}} \, ,
\ee
otherwise the particle decay would lead to excessive
cosmological entropy production which would
destroy the successful predictions of cosmological
nucleosynthesis. The constraint yields the following
bounds on the scale of supersymmetry breaking scale and 
on the gaugino condensation scale,
\be 
M_s \, > \,  10^{11} {\rm{GeV}} \ , ~~~~~~ \mu \, > \, 10^{14} {\rm{GeV}} \ .
\ee

Let us now study these constraints in our model. We
first need to find the location of the local minimum
of the potential $V(\Phi)$ given in (\ref{eq:scalarPotentil}).
The potential tends to zero at large negative values of
$\Phi$ and diverges for large positive values. The slope
of the potential in the limit $\Phi \rightarrow - \infty$ is
positive. Hence, the first extremum of the potential
is a local maximum, and the second is a local minimum.
Under the hypothesis that $\Phi_o$ is negative, then
the approximate form of the derivative of the potential is
\be \label{deriv}
\frac{\partial V}{\partial \Phi} \, \simeq \, V_0 
\bigl[ \frac{C^2}{4} e^{\Phi} - 2 a_0^2 A C \epsilon e^{- \Phi}
+ 4 A^2 a_0^3 e^{- 2 \Phi} \epsilon^2 \bigr] \, , 
\ee
with
\be
V_0 \, \equiv \, \frac{M_{4}^4}{4} b^{-6} 
\ee
and
\be \label{eps}
\epsilon \, \equiv \, e^{- a_0 e^{- \Phi}} \, .
\ee
Given the hypothesis that $\Phi_0$ is negative, and given that
$a_0$ is positive, $\epsilon$ is indeed a small quantity. This
explains the suggestive notation.

The local maximum of $V$ is given by the zero of (\ref{deriv})
obtained by balancing the first and the second term, the local
minimum $\Phi_0$ which we are looking for is given by
balancing the second and third terms. This yields
\be \label{inter}
e^{- \Phi} \epsilon \, = \, \frac{C}{2 A a_0} \, .
\ee
Making use once again of the hypothesis that $\Phi_0$ is
negative, we can simplify (\ref{inter}) to obtain
\be
\Phi_0 \, \simeq \, 
- {\rm{log}} \bigl[ \frac{1}{2 a_0} {\rm{log}} \bigl( \frac{2 a_0 A}{C} \bigr) \bigr] \, .
\ee
Thus, the self-consistency condition for $\Phi_0$ to be negative is
\be \label{cond4}
{\rm{log}} \bigl( \frac{2 a_0 A}{C} \bigr) \, > \, 2 a_0 \, .
\ee

Comparing (\ref{cond3}) with the condition (\ref{cond4})
required to have small string coupling, we see that it is
possible to satisfy both simultaneously, provided that
$C$ is a small number. Ways to achieve this were discussed
in \cite{Gukov}.

The dilaton mass can be obtained from the second
derivative of the potential $V(\Phi)$ at the minimum
$\Phi = \Phi_0$. The result is
\be
m_{\Phi}^2 \, = \, \frac{M_{4}^2}{8} C^2 
\frac{1}{2 a_0} {\rm{log}} \bigl( \frac{2 a_0 A}{C} \bigr) \, .
\ee
Thus, we see that the dilaton mass is suppressed compared to the
Planck mass by the constant $C$ which is required to be 
much smaller than $1$ in order to ensure weak string
coupling.

{F}rom Eq. \ref{eq:mu-g-relation} 
the scale of the gaugino
condensate is given by
\be
\mu \, \sim \, M_{4} \epsilon^{ \frac{1}{2 a_0 b_0}}  
\sim \, M_{4} \left( 2a_0 e^{-\Phi} \epsilon   \right)^{1/3}\, ,
\ee
where we have used the definition of $\epsilon$ from (\ref{eps}).
From (\ref{inter}) we then see that $\mu$ is suppressed
compared to the Planck scale by powers of $\frac{C}{ A }$, more
precisely
\be
\mu \, \sim \, M_{4} \bigl( \frac{C}{ A } \bigr)^{1/3} \, .
\ee
The result depends on (\ref{inter}), which requires 
$2 a_0 S \ll 1$. If this condition is not met then
\be 
\mu \, \sim \, M_{4} \left(  C-A e^{-a_0 S}  \right)^{1/3}\, .
\ee
Clearly, however, a high scale of gaugino condensation and hence
also of supersymmetry breaking appears more natural than a
low scale. Hence, the phenomenological constaints on the gaugino
mass are easily satisfied.

\section{Conclusions and Discussion}

We have shown that the gaugino condensation mechanism introduced
in \cite{Danos:2008pv} to stabilize the dilaton also leads to
supersymmetry breaking. The typical scale of supersymmetry
breaking is high. String gas cosmology (in the context
of perturbative heterotic string theory) thus provides a natural
way to stabilize all of the moduli fields and at the same time leads
to a non-supersymmetric low energy field theory.

We wish to emphasize that the stabilization of shape and size
moduli of the extra spatial dimensions is completely natural
in string gas cosmology. It relies on the effects of string modes which
carry both momentum and winding. These modes are not seen
in an effective field theory approach to string cosmology. A single
non-perturbative mechanism - namely gaugino condensation -
is sufficient to provide both dilaton stabilization and supersymmetry
breaking.

At this stage of the analysis, it appears that a high scale of supersymmetry
breaking is favored. Thus, it does not appear that a natural solution of
the hierarchy problem emerges. However, it would be of great interest
to extend our analysis to more realistic compactifications. With more
realistic toroidal orbifold compactifications it would then also be 
possible to obtain low energy field theories close to the Standard Model
(see e.g. \cite{orbifold} for a selection of references connecting
the Standard Model to heterotic string models).

\begin{acknowledgments}

We wish to thank Keshav Dasgupta, Andrew Frey, and Bret Underwood
for stimulating discussions. One of us (RB.) wishes to thank Joe
Polchinski for raising probing questions which motivated this
work. We thank Andrew Frey and Scott Watson for comments on the draft.
The research of R.B. is supported in part by an NSERC
Discovery Grant at McGill and by funds from the Canada Research
Chairs program.  The visit of S.M. to McGill University was supported by
a Commonwealth Scholarship for which we are grateful. W.X. is supported
in part by a Schulich Fellowship. U.Y. thanks the members of the 
McGill High Energy
Theory group for hospitality and financial support during a sabbatical visit.

\end{acknowledgments}

\section{Appendix}

\subsection{Radion}

The Einstein frame scale factor $b_E$ of an extra spatial dimension
becomes a radion field $\sigma$ in the four-dimensional
effective action. This field can be defined by
\begin{equation}
\sigma \, = \, \sigma_o ln\frac{b_E}{b_s} \, ,
\end{equation}
where $\sigma_o$ is a constant with dimension of mass. The
constant $b_s$ corresponds to the radius of an extra dimension
at the string scale. If moduli stabilization occurs at the
string scale (as in string gas cosmology in the absence
of a chemical potential for winding number) then $b_E = b_s$
and hence $\sigma = 0$.

\subsection{Gravitino mass and supersymmetry breaking}

The gravitino mass $m_{3/2}$ induced by the 
gaugino condensation mechanism
is given in terms of the scale $\mu$ by 
\begin{equation}
 m_{3/2} \, \sim \, \frac{\mu^3}{M_{4}^2} \, .
\end{equation}
And from the superpotential, the mass of the 
gravitino can be expressed as,
\be 
m_{3/2} \sim \frac{M_s^2}{M_{4}} \sim \frac{W}{M_{4}^2} \sim M_{4} \times \left(  C-A \ e^{-a_0 S} \right) \sim 2 M_{4}   a_0 e^{-\Phi} \epsilon
\label{gravitinomass}
\ee

The supersymmetry breaking scale $M_s$ is in turn given by
\be
M_s^2 \, \sim \, \frac{\mu^3}{M_{4}} \, .
\ee
Hence, for the gravitino mass to be order of TeV the supersymmetry 
breaking scale must be about $10^{14}$GeV. This constrains the 
value of $e^{\varphi}$ in Eq.(\ref{eq:mu-phi-relation}).

\end{document}